\documentclass{format/sig-alternate-10pt}

\usepackage[utf8]{inputenc}
\usepackage{xcolor}
\usepackage{indentfirst}
\usepackage{url}
\usepackage{graphicx}
\usepackage{float}
\usepackage{subcaption}
\usepackage{enumitem}
\usepackage{array}
\usepackage[linesnumbered,lined,boxed,commentsnumbered]{algorithm2e}

\definecolor{darkgreen}{RGB}{47,109,79}
\definecolor{darkblue}{RGB}{57,79,99}
\usepackage[bookmarks=false, colorlinks=true, plainpages=false, linkcolor=darkgreen, citecolor=darkgreen, urlcolor=darkblue, filecolor=blue]{hyperref}
\usepackage[auth-lg]{authblk}

\title{There goes Wally: Anonymously sharing your location gives you away}

\author[1]{Apostolos Pyrgelis}
\author[2]{Nicolas Kourtellis}
\author[2]{Ilias Leontiadis}
\author[2]{Joan Serr\`a}
\author[3]{Claudio Soriente}
\affil[1]{University College London, apostolos.pyrgelis.14@ucl.ac.uk}
\affil[2]{Telefonica Research Barcelona, firstname.lastname@telefonica.com}
\affil[3]{NEC Labs Europe, claudio.soriente@emea.nec.com}

\date{\today}

\newcommand{\descr}[1]{\vspace{0.05cm} \noindent \textbf{#1}}

\begin{document}
\sloppy

\maketitle

\thispagestyle{empty}
\pagestyle{empty}

\begin{abstract}
With current technology, a number of entities have access to user mobility traces at different levels of spatio-temporal granularity. At the same time, users frequently reveal their location through different means, including geo-tagged social media posts and mobile app usage. Such leaks are often bound to a pseudonym or a fake identity in an attempt to preserve one's privacy. In this work, we investigate how large-scale mobility traces can de-anonymize anonymous location leaks. By mining the country-wide mobility traces of tens of millions of users, we aim to understand how many location leaks are required to uniquely match a trace, how spatio-temporal obfuscation decreases the matching quality, and how the location popularity and time of the leak influence de-anonymization. We also study the mobility characteristics of those individuals whose anonymous leaks are more prone to identification. Finally, by extending our matching methodology to full traces, we show how large-scale human mobility is highly unique. Our quantitative results have implications for the privacy of users' traces, and may serve as a guideline for future policies regarding the management and publication of mobility data.
\end{abstract}

\section{Introduction}
\label{sec:intro}

With the proliferation of mobile devices and the wide deployment of sensing infrastructures, an increasing number of entities have access to the mobility patterns of their customers. Mobile OS developers, such as Google or Apple, collect mobility data to improve their services or to enhance their revenues. Similarly, mobile apps such as fitness trackers might collect location information and even share it with third-party libraries (e.g., advertisers). Mobile connectivity providers need to maintain the location of their subscribers to route calls/data and to aid emergency response services. Furthermore, insurance companies harvest location data from on-board units available on vehicles to tailor their policies. Finally, some governments have deployed sensing infrastructure that can record mobility traces (e.g.,~road-side cameras to monitor and regulate vehicles). These are just some examples of entities that can passively collect location information with different levels of spatio-temporal granularity.

At the same time, users often reveal information about their whereabouts, either voluntarily or without even realizing it. Location-based online social networks such as Foursquare allow users to check-in at various places and publicize this information for other users to see. Online review systems like TripAdvisor let users review services or facilities at precise locations. Mainstream social networks such as Facebook and Twitter allow users to geo-tag their posts. Moreover, large amounts of photos are uploaded daily that either contain location meta-data or depict landmarks that can be easily associated to a given location. Finally, users might involuntary leak coarse-grained location information to advertisement networks and service providers each time they visit a web page in the form of an IP address that can be geo-localised with an error of a few tens of kilometers.

The difference between ``active'' location leaks and passively-collected location data is mainly three-fold. First, they can differ in \emph{temporal resolution}. Active leaks are normally sporadic, as they happen when, for instance, the user posts geo-tagged content on a social network. Passive data collection can happen at a much higher temporal resolution, depending on the sampling rate of each entity (e.g., mobile network provider, fitness tracker, OS, mobile app). Second, they can differ in \emph{spatial granularity}. Active leaks from IP geo-location services provide granularity down to city level (i.e.,~tens of kilometers), whereas GPS-tagged posts can go down to a few meters. Similarly, passively collected data is highly dependent on the means of the data collection technology. Third, they can differ in the \emph{identity} associated with the data. Passively-collected location data is usually associated to the real user identity for, e.g., billing purposes. Active location leaks can be bound to pseudonyms that users adopt in order to achieve some degree of anonymity~\cite{freudiger11fc,kang13chi,peddinti17sp}. In such a scenario, users expect that the published content (and location) cannot be traced back to their real identity nor to the other places they have visited. While such privacy guarantees may hold on a single platform, it is unclear what could happen if two or more platforms cross their datasets in order to identify a user. We argue that de-anonymization of anonymous location leaks is a real threat given the number of entities that keep identified location traces of their customers or users.

To assess the magnitude of this threat, we leverage an unprecedented scale, country-wide dataset of network events from an European mobile operator, and use it as a mobility oracle. We opt not to use a second dataset of anonymized location leaks but rather emulate them. The motivation behind this choice is two-fold. On the one hand, matching an identified set of traces against a set of anonymous leaks may lead to false positives when the real owner of the leak is not included in the dataset (i.e.,~a user of another mobile operator). On the other hand, we cannot attempt to recover the real identity of the author of an anonymous post, as this may lead to ethical issues and would require explicit user consent. We therefore use only the mobility dataset and reason about (1) the likelihood of correctly matching an anonymized location leak to one of the traces of the dataset, (2) what makes a user more identifiable than others, and (3) the overall uniqueness of large-scale mobility traces.

Our measurement study demonstrates that just 3--4 anonymous location leaks within the time frame of one day are sufficient to uniquely identify one's mobility trace among tens of millions. As we examine why some users are more identifiable than others, we observe that exposing less popular locations eases the performance of matching, and that revealing one's location during a day's working hours contributes more to the identification probability. Moreover, we find that highly mobile users are more prone to identification. Our results show how spatio-temporal obfuscation applied to mobility traces can decrease the possibility of such identification. In fact, we show that a 12$\times$ decrease in the temporal granularity, or a 25$\times$ decrease in the spatial dimension brings up to 4$\times$ reduction in the probability of matching one's identity. While these results can potentially be used as guidelines for the privacy-respecting storage and publication of such traces, we argue that spatio-temporal obfuscation alone is not sufficient to prevent the privacy threat represented by the uniqueness of large-scale mobility traces.

The rest of the paper is organized as follows. Section~\ref{sec:related} surveys related work and points out the differences with ours. Section~\ref{sec:methodology} introduces basic notation and the methodology we use to analyze mobility traces, while Section~\ref{sec:dataset} presents our dataset. We provide quantitative experimental results in Section~\ref{sec:experiments} and discuss their implications in Section~\ref{sec:discussion}. Finally, Section~\ref{sec:conclusion} provides some concluding remarks.
\section{Related Work}
\label{sec:related}

A number of research studies have demonstrated how the inherent nature of location data, as well as the patterns hidden in it, can harm user privacy. For example, Golle and Partridge~\cite{golle2009anonymity} use census data to demonstrate the uniqueness of home/work location pairs across a fraction of the US population. De Montjoye et al.~\cite{de2013unique} use the Call Detail Records (CDRs) of 1.5\,M users over a 15-month period and perform a measurement study to understand the uniqueness of human mobility, as well as its relation to the spatial and temporal resolution of the traces. Zang and Bolot~\cite{zang2011anonymization} use anonymized CDRs and examine the uniqueness of the ``top N'' locations, with the rationale that an adversary may have available identified data regarding the top N locations of a population (i.e., census data) and could use it to de-anonymize the CDRs. Rossi et al.~\cite{rossi2015spatio} use the GPS trajectories of taxis (536~vehicles) and mobile phone users (182~individuals) to show that mobility features like speed, direction, and travel distance can be exploited to map trajectories to identities.

Another parallel line of work is focused on exploiting the characteristics of human mobility in order to link users across different data sources and de-anonymize them. Srivatsa et al.~\cite{srivatsa12ccs} de-anonymize mobility traces using social networks as a side-channel. They show how encounters in mobility traces can be mapped to relationships in a social network graph for three datasets of less than 100 users each. Freudiger et al.~\cite{freudiger11fc} study the dynamics of anonymous location leaks that can erode user privacy. They use two datasets (143~users over a year, and 200~users over two years, respectively) and try to profile users by identifying points of interest such as work and home locations. De Mulder et al.~\cite{de2008identification} propose statistical techniques to first create mobility profiles of users of GSM networks, and subsequently match those profiles against anonymized location data. The techniques they employ are evaluated with a dataset that spans 100~users ``followed'' across nine months. Ma et al.~\cite{ma2013privacy} use various estimators to calculate the similarity between mobility traces of roughly 7\,k cabs and buses, aiming to de-anonymize users, while Gambs et al.~\cite{gambs2014anonymization} show that mobility traces can be exploited as a signature to identify an individual in a set of anonymized traces. They leverage a number of datasets, ranging from one with 5~researchers to another with 500~taxi drivers.

Along the same line, Cao et al.~\cite{cao2016automatic} demonstrate how to match mobility traces derived from different data sources to the same identity. Their dataset includes roughly 32\,k users and spans across seven months. Similarly, Riederer et al.~\cite{riederer2016linking} propose a self-tunable algorithm that determines the most likely matching between users of two location-based datasets, validating it on datasets of a few thousand users. Naini et al.~\cite{naini2016matchingstatistics} introduce techniques for matching users between GPS trajectory datasets which exploit their mobility histograms as fingerprints. Kondor et al.~\cite{kondor2017towards} study the problem of matching mobility datasets on the scale of an urban setting and propose an efficient spatio-temporal search algorithm for that. They evaluate their algorithm by matching a mobile communication dataset of 2.8\,M users with a transportation one of 3.3\,M users, both corresponding to the city of Singapore. However, due to the absence of ground truth, their results only capture an estimate of matchability between the two datasets.

More recently, Wang et al.~\cite{wangcikm2017} use mobility pattern mining to link user identifiers across an HTTP dataset obtained by an internet service provider and three external datasets capturing instant messaging, electronic commerce, and social networking transactions of a subset of the users. In a follow up work~\cite{wangndss2018}, they evaluate the performance of various mobility trajectory de-anonymization algorithms. They experiment on a large-scale dataset containing the mobility trajectories of 2.1\,M users of a Chinese mobile network, and two external datasets corresponding to a subset of the same user population as obtained from a social network (Weibo) and a check-in service (Dianping). The results demonstrate that existing methods under-estimate spatio-temporal mismatches as well as the noise in the data generated from various sources, and the authors propose novel algorithms to account for both.

Compared to the works mentioned above, our study differs in the nature and size of the dataset, as well as in the research questions it answers. First, we concentrate on one day-worth of data, and show that user anonymity can be easily infringed with a comprehensive dataset of such a short time-span. Second, our dataset accounts for tens of millions of customers. Therefore, it features at least an order of magnitude more users compared to the datasets used in prior studies. Third, mobility traces are generated from mobile network events. While these might have a coarser spatio-temporal granularity than, for instance, GPS traces, they scale well to country-level regions and population. Importantly, mobile network events have a much finer granularity compared to trajectories derived by, for instance, CDRs, which can feature more than two orders of magnitude less events. Fourth, the size of the country we monitor differs from the one monitored in previous works. With the exception of Zang et al.~\cite{zang2011anonymization}, which consider the entire US, our study leverages mobility traces from a much larger geographical region than existing works (i.e.,~it features the diversity of mobility at various geographical scales). Finally, while our study evaluates the uniqueness of mobility traces, we also investigate the characteristics that make them prone to identification by examining, among others, location popularity, time of day, and user mobility characteristics.
\section{Methodology}
\label{sec:methodology}

\subsection{Notation}
\label{sec:notation}

We consider a set of users $U=\{ u_i\}$ that move among a set of locations $S=\{s_i\}$ during the time instances of a set $T=\{t_i\}$. Each location\footnote{We drop the subindices where there is no ambiguity.} $s=(x,y)$ is represented by a pair of grid coordinates $x$ and $y$ whose spatial granularity is controlled by the parameter $\Delta_{xy}$. For instance, if $\Delta_{xy}=100$\,m, then each $s$ covers an area of 100$\times$100=10,000\,m$^2$. The set $T$ represents the time frame in which locations are reported, and its granularity is controlled by the parameter $\Delta_t$. For instance, if $\Delta_t=1$\,h, then $|T|=24$\,h for the time frame of a single day. The sets $S$ and $T$ are dependent on the discretization applied to the continuous spatial and temporal dimensions, respectively.

The mobility trace of a user is computed from the set of \textit{events} that she generates. An event is a triplet of the form $(u, s, t)$, which indicates that user $u$ is in location $s$, at time $t$. Note that, depending on the temporal granularity $\Delta_{t}$, a user might generate multiple events at a specific time $t$. Next, we define a \textit{mobility trace} of a user $u$ as a set of tuples $M_u=\{ (s_i,t_j) \}$, where $s_i\in S$ and $t_j\in T$. Similarly, we define an \textit{anonymous} spatio-temporal leak of some unknown user as the set of tuples $L=\{ (s_i,t_j) \}$, $s_i\in S$ and $t_j\in T$. We use $k$ to denote the number of events (tuples) in $L$, that is, $k=|L|$. Finally, we denote a dataset of user mobility traces as $D=\{M_{u_i}\}$, where $u_i\in U$.
 
\subsection{Matching Spatio-Temporal Leaks}
\label{sec:traject_match}

The main purpose of this study is to estimate the number of events required to uniquely identify a user and her entire mobility trace. To this end, we follow a matching approach that estimates the probability of an anonymous spatio-temporal leak $L$ leading to a unique trace in the dataset $D$. Given $L$ and a user mobility trace $M_u$, we define a \textit{match} between them if all the tuples in $L$ can also be found in $M_u$. That is 
\begin{equation}
    \mu\left(L,M_u\right) = \delta\left( \left| L \cap M_u \right| , \left| L \right| \right) ,
    \label{eq:match}
\end{equation}
where $\delta$ is the Kronecker delta function\footnote{For two variables $i$ and $j$, $\delta(i, j)=1 \,\, \text{if} \,\, i=j, \, 0 \, \text{otherwise}$.} and $\cap$ denotes tuple set intersection\footnote{A tuple is considered to be equal to another if all elements are respectively equal: $(a,b)=(c,d)$ if $a=c$ and $b=d$.}. We denote the total number of matches for the anonymous spatio-temporal leak $L$ in the set $D$ as
\begin{equation}
    \nu\left(L,D\right) = \sum_{M_u\in D} \mu\left(L,M_u\right) .
    \label{eq:num_matches}
\end{equation}
A \textit{unique match} occurs when a single mobility trace $M_u$ in the entire set $D$ matches the spatio-temporal leak $L$. We depict this by
\begin{equation}
    \xi\left(L,D\right) = \delta \left( \nu\left(L,D\right) , 1 \right) .
    \label{eq:unique_match}
\end{equation}

To estimate the probability of an anonymous spatio-temporal leak $L$ (with $k$ events) identifying a unique trace, we sample from $D$ a set of spatio-temporal leaks $Z=\{L_u\}$, $u\in U$. We do so by selecting a user $u$ and choosing $k$ events from her mobility trace $M_u$ (user selection and event choices are performed uniformly at random without replacement). Finally, we perform matching for all the leaks $L_u \in Z$, and we estimate the probability that a leak with $k$ events yields a unique match as
\begin{equation}
    \rho_k\left(Z,D\right) = \frac{1}{\left| Z \right|} \sum_{L_u \in Z} \xi\left(L_u,D\right) .
    \label{eq:prob_unique_match}
\end{equation}

\subsection{Spatial Bounding Box}
\label{sec:boundingbox}

To speed up the calculation of Eq.~\ref{eq:num_matches} and reduce the number of comparisons that we perform when we match a spatio-temporal leak to a set of mobility traces, we employ the concept of spatial bounding box. The bounding box of a mobility trace is a spatial box that captures the geographic area in which the trace owner is moving. It is defined by the minimum and maximum coordinates computed on all the locations of the trace. More precisely, for a mobility trace $M_u$, we define its bounding box $b_{u}$ as a tuple such that $\forall ~s=(x,y)\in M_u$,
\begin{equation*}
    b_{u} = \left( x_{\text{min}},y_{\text{min}},x_{\text{max}}+\Delta_{xy},y_{\text{max}}+\Delta_{xy} \right) ,
\end{equation*}
where $x_{\text{min}}$ and $x_{\text{max}}$ respectively correspond to the minimum and maximum values of the $x$ coordinate contained in the trace (and analogously for $y$ coordinate). Note: we can also define a bounding box $b_{L}$ for an anonymous spatio-temporal leak $L$.

Given two bounding boxes $b_L$ and $b_u$, we can compute their overlap, which corresponds to their normalized intersection:
\begin{equation*}
    o\left(b_L,b_u\right) = \frac{ \alpha(\iota(b_L,b_u)) }{ \min\left( \alpha(b_L) , \alpha(b_u) \right) } ,
\end{equation*}
where $\iota$ denotes the spatial intersection of two bounding boxes (which is another bounding box) and $\alpha$ indicates the area of a bounding box. It is then safe to skip the computation of Eq.~\ref{eq:num_matches} for $L$ and $M_u$ when $o(b_L,b_u)=0$ (that is, when the bounding box of a spatio-temporal leak does not intersect with that of a mobility trace).

\subsection{Matching Mobility Traces}
\label{sec:wholetraj}

Thus far, our methodology focused on estimating the probability that an anonymous spatio-temporal leak $L$ uniquely identifies a mobility trace $M_u$ (where $|L|<<|M_u|$). We now extend it to account for matching whole mobility traces between themselves, as one of our objectives is to also study their uniqueness. To do so, we follow the same matching approach as before (Section~\ref{sec:traject_match}), but now compare samples of mobility traces $M_u$ to all the other traces $M_{u'} \in D$. For that we use Eqs.~\ref{eq:match}--\ref{eq:prob_unique_match} to estimate the probability of a mobility trace yielding a unique match in the set $D$. To speed up the calculation of Eq.~\ref{eq:num_matches}, we here also employ the concept of bounding boxes explained above, and we compare two traces $M_u$ and $M_{u'}$ when their normalized bounding box intersection exceeds a threshold $r$, that is, if $o(b_u, b_{u'}) > r$.

We note however that the current approach to matching is very \textit{strict} when dealing with whole traces. Due to Eq.~\ref{eq:match}, two mobility traces $M_u$ and $M_{u'}$ will only match if their owners $u$ and $u'$ visited exactly the same locations at each instance of time frame $T$. Therefore, we shall consider the previous approach as an upper bound on the uniqueness of a trace. 

To contrast this upper bound, we also \textit{relax} the conditions on which two traces match, aiming to also estimate a lower bound on the probability of uniqueness. For that, we take into account characteristics of mobility traces that arise due to their temporal granularity $\Delta_t$. On the one hand, when $\Delta_t$ is small, we observe that mobility traces can be very sparse, as users might not generate events at a time $t \in T$. On the other hand, when $\Delta_t$ is large, we observe that mobility traces can be dense, as users might visit a number of locations within a time instance. To favor matches in these situations, we define that two traces $M_{u}$ and $M_{u'}$ match at a given $t\in T$ if (1) they intersect at least in one location or (2) one of them does not contain events for that time $t$. Then $\mu(M_u,M_{u'})=1$ if they match $\forall~t\in T$, and 0 otherwise.
\section{Dataset Description}
\label{sec:dataset}

\begin{table}[t]
    \center
    \resizebox{0.85\linewidth}{!}{
    \begin{tabular}{| l || c |}
      \hline 
      \textbf{Feature} & \textbf{Average (Std.~Dev.)} \\
      \hline \hline
      \textbf{Spatial coverage} & Whole country \\ \hline
      \textbf{Users} & More than 30\,M  \\ \hline
      \textbf{Sectors} & More than 150\,k  \\ \hline
      \textbf{Unique locations} & More than 40\,k \\ \hline 
      \textbf{Users per location per day} & More than 3\,k \\ \hline 
      \textbf{Events per user per day} &  279 (506) \\ \hline 
      \textbf{Unique locations per user}  & 15 (23) \\ 
      \hline
    \end{tabular}
    }
    \caption{Description of the dataset used. Note: the exact numbers cannot be shown at the request of the mobile operator.}
    \label{table:dataset}
\end{table}

Our dataset consists of anonymous traces collected from a large European cellular network provider that is serving tens of millions of mobile subscribers. Each trace is a time series of \emph{mobile events} as defined below. When compared to traces based on CDRs, ours provide significantly finer sampling of the mobility trace. In particular, we observe that calls and SMS correspond to around 1\% of the number network events generated by users on a daily basis.

\subsection{Mobile Events}

We tap into the mobility management entity (MME) of the network provider. The MME is a key component in the cellular network, which is responsible for handling ``control plane'' messages related to paging, radio channel requests, and handovers, in order to route calls and data to and from subscribers as they move from one base station to another (BTS/NodeB/eNodeB, depending on the technology). Compared to CDRs, where location information is only collected when a subscriber initiates/receives a call or an SMS, the MME maintains a significantly richer view of the location of the subscribers. In particular, a mobile event is generated every time a mobile device:
\begin{itemize}
    \item Activates/deactivates in the network (i.e., when the user switches on and off her phone)
    \item Makes/receives a call or sends/receives an SMS (i.e., what is included in CDRs)
    \item Moves from one location area code to another (i.e., the so-called handovers)
    \item Changes from one technology to another (i.e., between 2G, 3G, and 4G)
    \item Requests access to data (2G/3G) or requests a high-speed data channel (4G)
    \item Is actively pinged by the network if no other event is registered for 2 hours (i.e., a timeout to check if the devise is still alive)
\end{itemize}

Overall, the MME handles several hundreds of events per device per day, almost two orders of magnitude higher than traditional CDRs.
These events contain the encrypted user identifier (rotated daily), a timestamp, and the location of the associated base station used to deliver service to the user. This implies that the exact location of a subscriber is not known. The MME only registers the location of the base station (i.e.,~antenna) to which the device is connected. Base stations have varied coverage, ranging from less than 150~meters to tens of kilometers depending on the deployment density and the radio propagation characteristics (such as obstacles, hills, or mountains). Typically, short-range base stations are used in densely populated areas (to share the load), whereas long-range ones are used in rural areas. This means that the expected user displacement in urban areas is smaller than that in rural areas, and can reach as low as 70~meters~\cite{cell-star}.

Overall, there are more than 40\,k distinct locations in the country of study, with each location serving a wide number of subscribers per day. For example, sites that cover highly dynamic locations (highways, airports, train stations) might serve hundreds of thousands of users daily for small periods of time, whereas sites that serve residential and rural areas might only provide connectivity to a few thousand subscribers. On average, each location registers less than 10\,k users over the course of a whole day.

\subsection{Dataset Pre-processing}

Note that a \emph{location} might be covered by multiple base stations. For example, three directional (60 degree) panels might be used to cover the area around an antenna and, furthermore, there can be multiple installations per technology (2G/3G/4G). In this work, we combine all the base stations in the same location with a single identifier that is represented by its latitude and longitude. Furthermore, to avoid any ping-pongs~\cite{ekiz2005overview} from nearby antennas that cover the same location (e.g., installations within a mall) we combine together events that correspond to sites that are less than 150\,m apart. The result is a mobility trace that can be expressed as a time series of \textit{events}, represented by triplets of the form (\textit{timestamp}, \textit{latitude}, \textit{longitude}). In our dataset, each trace contains an average of 279~events and each user reports an average of 15~unique locations during the day (Table~\ref{table:dataset}).

\subsection{Note on Customer Privacy}

The dataset is anonymized by the provider and only temporarily stored for operating purposes. Furthermore, the computing infrastructure does not allow us to extract mobility traces for specific individuals, but only high-level aggregates or results of algorithms that combine together information from multiple users, like we do in this study. Also note that, by request of the provider, we cannot disclose full information on the dataset such as the exact number of subscribers, number or location of antennas, etc. However, we provide here the order of magnitude (Table~\ref{table:dataset}).
\section{Experimental Results}
\label{sec:experiments}

With our measurement study, we aim to answer the following research questions:
\begin{enumerate}
    \item How many anonymous location leaks does it take to identify their owner, using mobility traces only?
    \item Which spatio-temporal granularity brings sufficient obfuscation to those traces?
    \item How does the probability of identification associate with the popularity of the locations leaked?
    \item How does the probability of identification associate with the time of day the leaks happen?
    \item How does the probability of identification associate with the mobility characteristics of users?
    \item How unique is a user's mobility trace?
\end{enumerate}
The first five questions motivate us to study the various ways that the probability of de-anonymization of users is associated with different factors such as user characteristics, their mobility behavior through the day, the characteristics of locations visited, etc.
The final question motivates us to study the uniqueness that users have with respect to their mobility behavior in space and time.
In the next paragraphs, we first describe our experimental setup, followed by results providing answers for each of these questions.

\subsection{Experimental Setup}

\descr{Location Leaks.}
We generate a set of anonymous location leaks per user by selecting uniformly at random $k$ points from her mobility trace. Therefore, $k$ is a parameter which represents the number of \textit{leaks} that users may have exposed by, e.g.,~geo-located social media posts. We experiment with $k \in \{1, 2, \dots 10 \}$.

\descr{Exhaustive Search.}
All experiments performed require pairwise comparisons between users' mobility traces $M_u$ in order to assess the probability of matching and uniqueness of those traces. Due to high computation complexity, we cannot handle all pairwise comparisons (over $10^{15}$ comparisons).
Instead, we select a large random sample of traces and use those to generate anonymous location leaks. The sample is large enough to give us statistically significant results for the questions studied, and it is compared against \emph{all} available traces in $D$.

\descr{Spatio-Temporal Resolution.}
We study the effects of temporal and spatial resolution on the probability of de-anonymizing a user by pre-processing the events contained in $M_u$ using specific temporal and spatial granularity for data aggregation, as explained in Section~\ref{sec:methodology}. In particular, assuming that users' mobility is typically governed by the space dimension, we vary the spatial granularity $\Delta_{xy}\in\{ 0.2, 1, 5, 25, 125 \}$ km. We assume that such values represent typical scenarios of user movement in the range of the same building block, neighborhood, city, county, and country, respectively. Under a similar premise, we vary the temporal resolution $\Delta_t\in\{ 5, 15, 30, 60 \}$ minutes to represent such movements assumed earlier.

\descr{Bounding Box Setting.}
As mentioned, our method uses bounding boxes to describe mobility traces and to reduce computation when comparing traces. For matching leaks to mobility traces, we discard the leak-user pairs that do not overlap, i.e.,~$o(b_L,b_u)=0$ (Section~\ref{sec:boundingbox}). However, in the particular case of matching mobility traces, we provide bounds to the probability of a unique match by setting a minimum overlap threshold $o(b_L,b_u)>r$ (Section~\ref{sec:wholetraj}). We tested different values for $r\in\{ 0.001, 0.01, 0.1, 0.5, 0.9\}$, and found that with $r = 0.5$, we reduce the comparisons by approximately two orders of magnitude while still maintaining a good performance of the matching algorithm.

\begin{figure}[t]
\center
    \includegraphics[width=0.38\textwidth]{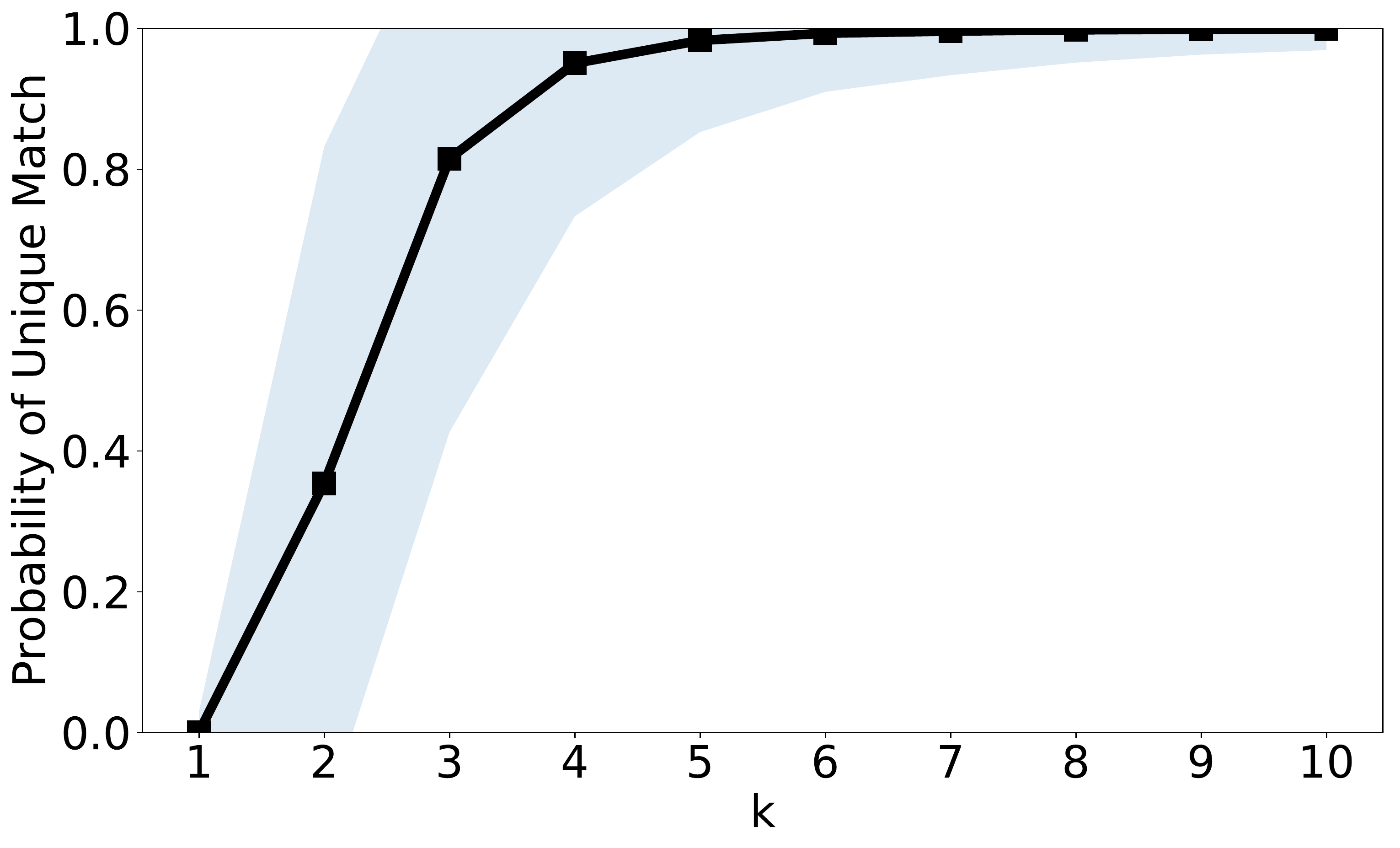}
    \caption{Probability of uniquely matching a mobility trace $\rho$ vs.~number of location leaks $k$. The experiment was conducted with $\Delta_t=5$\,min and $\Delta_{xy} = 0.2$\,km. The shaded area denotes one standard deviation.}
    \label{fig:q1}
\end{figure}

\begin{figure}[t]
\center
    \includegraphics[width=0.4\textwidth]{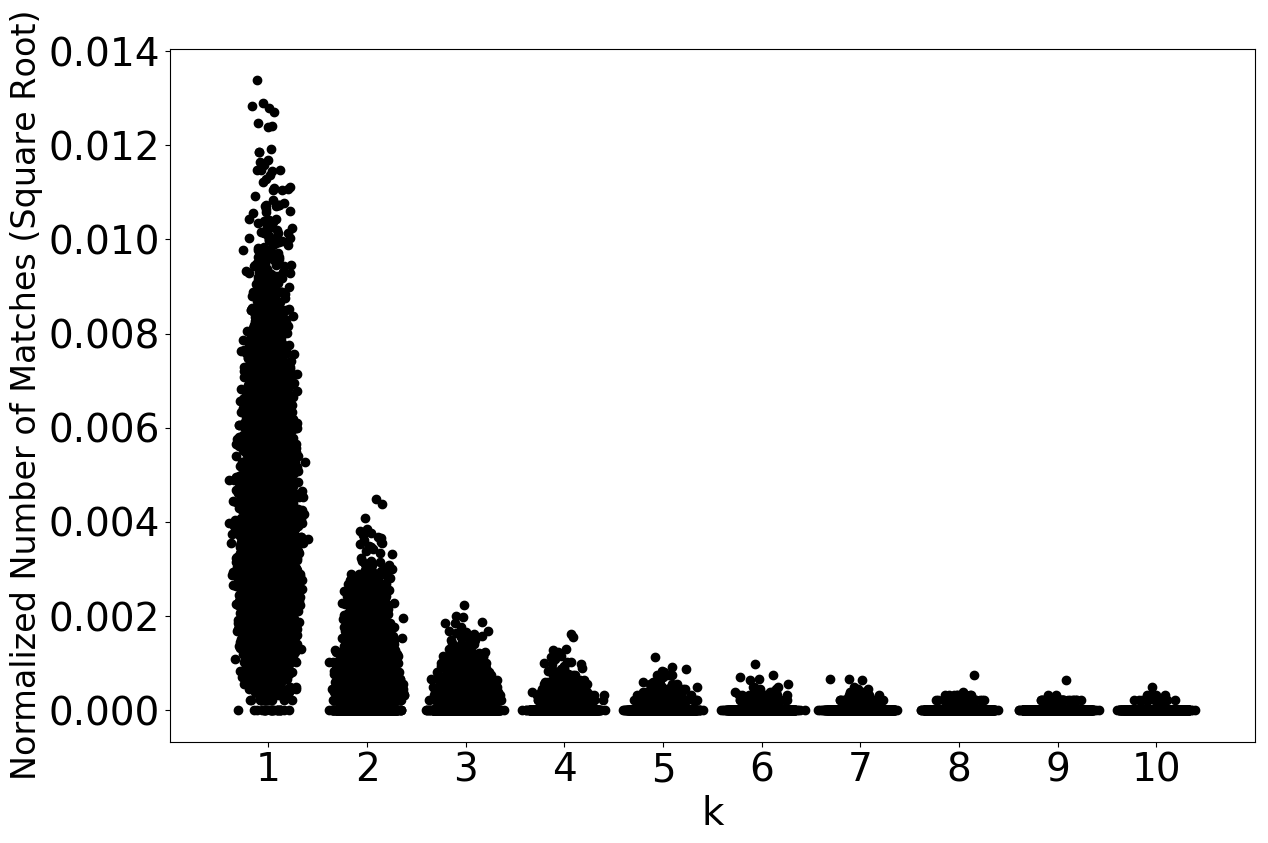}
    \caption{Square root of the normalized number of matches, $\sqrt{\nu}$, vs.~number of leaks $k$ (we take $\sqrt{\nu}$ for ease of visualization). Experiment conducted with $\Delta_t = 5$\,min and $\Delta_{xy} = 0.2$\,km.}
    \label{fig:q1-matches}
    \vspace{-0.4cm}
\end{figure}

\subsection{How many leaks are needed?}

\begin{figure*}[t]
\center
    \begin{subfigure}[b]{0.38\textwidth}
		\includegraphics[width=1.0\textwidth]{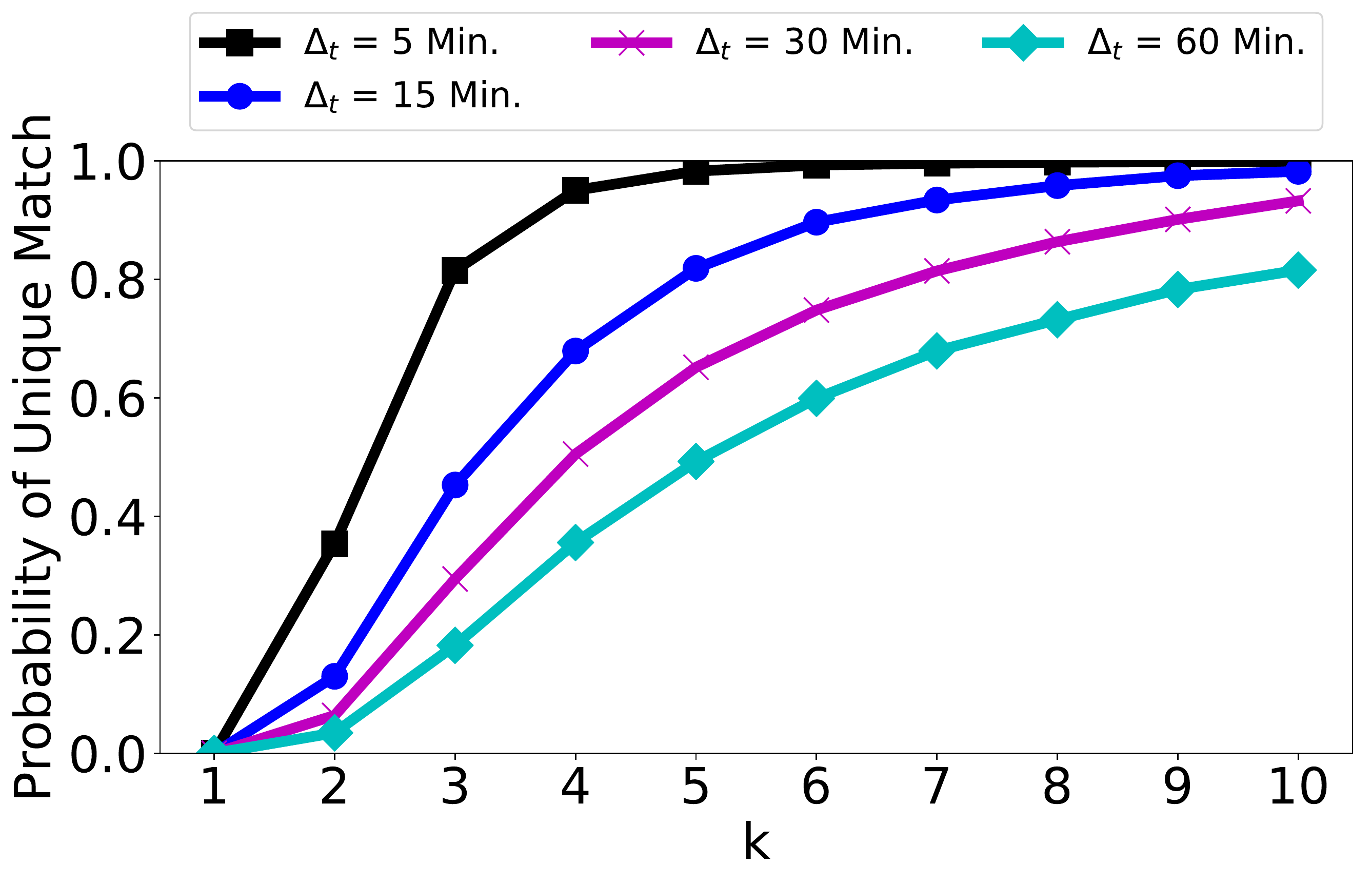}
		\vspace{-0.5cm}        				
		\caption{We keep constant $\Delta_{xy} = 0.2$\,km.}
		\label{fig:q2-varydt}
	\end{subfigure}
	~
    \begin{subfigure}[b]{0.38\textwidth}
		\includegraphics[width=1.0\textwidth]{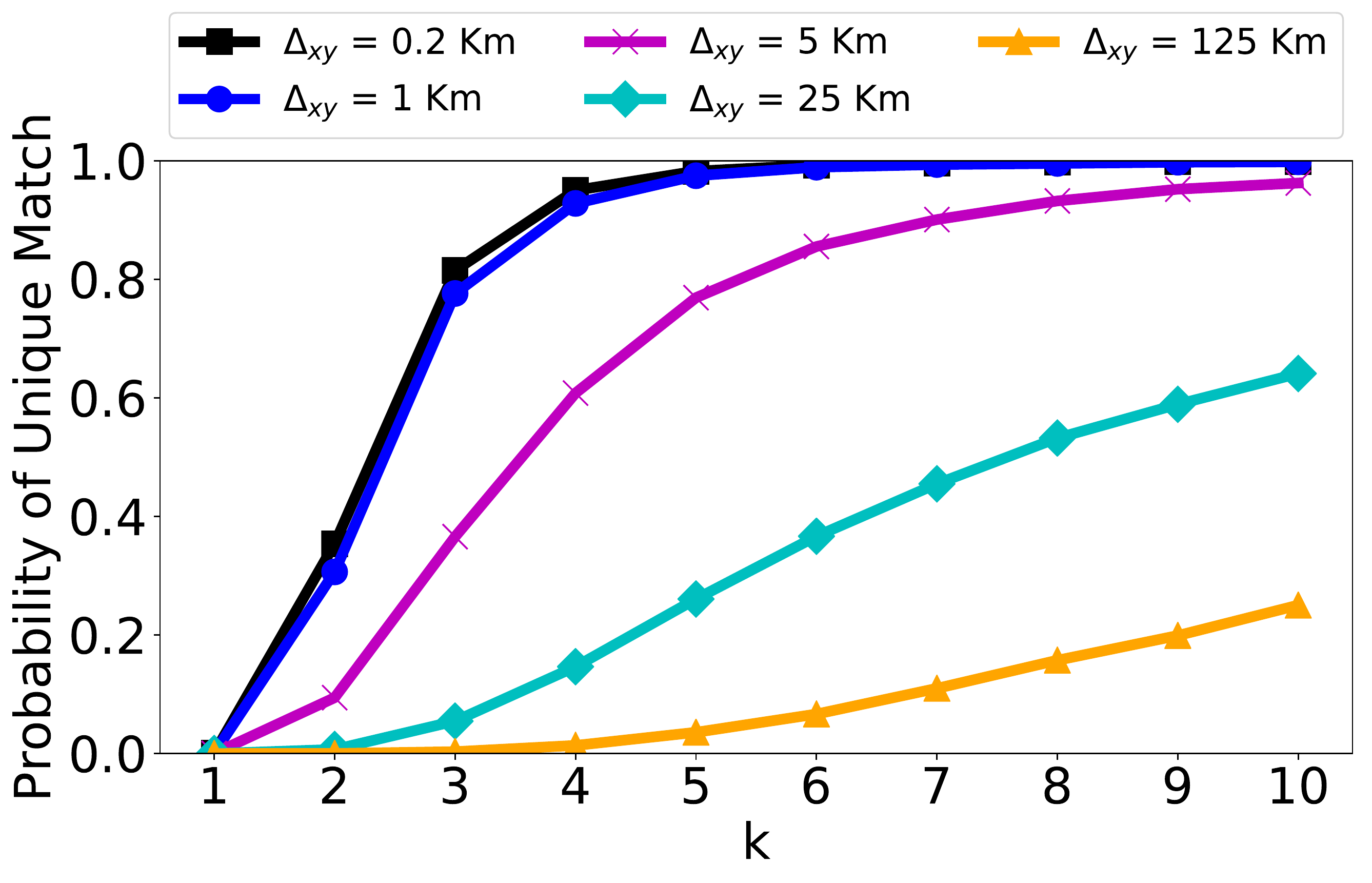}
		\vspace{-0.5cm}        
        \caption{We keep constant $\Delta_t = 5$ minutes.}
        \label{fig:q2-varydx}
    \end{subfigure}
    \caption{Probability of uniquely matching mobility trace $\rho$ vs.~number of leaks $k$ using variable (a) temporal, (b) spatial resolution.}
    \label{fig:q2}
    \vspace{-0.5cm}
\end{figure*}

Our first set of experiments aims to investigate how many anonymous location leaks $L$ from user $u$ are sufficient to match her to a mobility trace in our dataset $D$. We use a randomly selected set of 30\,k users, sample events from their traces to form $L$, and perform matching across all other traces in $D$. Figure~\ref{fig:q1} plots the average probability of uniquely matching a mobility trace based on the number of location leaks $k$ that are exposed from the user when we set the temporal granularity to 5\,minutes and the spatial one to 0.2\,km. While the probability of matching a trace is negligible when only one location is leaked, we observe that it increases to 0.35 and 0.8 when $k$ becomes 2 and 3, respectively. Interestingly, we note that 4 location leaks are sufficient to uniquely match 95\% of the traces in our dataset, indicating that large-scale collection of mobility data can seriously harm users' privacy. As the number of leaks increases further, the probability of matching a trace reaches 1 and the majority of the leaks are uniquely matched to traces (note how the standard deviation of the matching probability decreases as $k$ increases).

While Figure~\ref{fig:q1} displays the average probability of uniquely matching a trace, it does not capture the ``anonymity set'' of those leaks that cannot be uniquely matched to a mobility trace. To this end, Figure~\ref{fig:q1-matches} displays a scatter plot of the normalized number of matches for variable number of leaks, following our previous experimental setting with $\Delta_{t}=5$\,min and $\Delta_{xy}=0.2$\,km. Surprisingly, when one location is leaked, we observe that the normalized number of matches is already very small, with the median being approximately $10^{-5}$. This indicates that one leak is sufficient for our matching algorithm to exclude a very large number of traces. When two leaks from a mobility trace are exposed, we note an approximate decrease of two orders of magnitude in the number of matches, with a median around $10^{-7}$. As $k$ increases, i.e.,~as more spatio-temporal points are leaked, we see that more traces are matched uniquely, and the median becomes 0. In fact, with $k \geq 3$, our matching algorithm can uniquely identify all users besides some exceptional outlier traces, which is consistent with the probability of unique match shown in Fig.~\ref{fig:q1}.

\descr{Takeaways:}
\begin{itemize}
\item Just 3 (resp.~4) anonymous location leaks are sufficient to uniquely match 80\% (resp.~95\%) of users.
\item Exposing 2 leaks instead of 1 decreases by 100$\times$ the set of candidate matching traces.
\end{itemize}

\subsection{Does spatio-temporal granularity matter?}

Next, and using the same set of users selected earlier, we study how the temporal and spatial obfuscation of the events contained in a mobility trace affects the performance of our matching algorithm.

\descr{Temporal Resolution.}
To observe the effect of temporal granularity on matching anonymous location leaks to mobility traces, we fix $\Delta_{xy}=0.2$\,km and vary $\Delta_t$. Figure~\ref{fig:q2-varydt} displays the probability of uniquely matching a trace when $k$ leaks are revealed. Overall, we observe that obfuscating leaks in the time dimension reduces the probability of matching. For instance, while the probability of matching a mobility trace with $k=4$ is $0.95$ for $\Delta_t=5$\,min, it goes down to $0.67$ when $\Delta_t=15$, $0.5$ when $\Delta_t=30$, and $0.35$ when $\Delta_t=60$\,min. Moreover, we observe that as the temporal granularity decreases, more leaks are required to match a trace, indicating that such an obfuscation approach can bring some privacy protection to the traces. Nonetheless, we highlight that 10 leaks with a temporal resolution of 1\,h are still sufficient to match 81\% of traces, hinting to the high uniqueness of large-scale human mobility and the inherent difficulty in obfuscating mobility traces.

\descr{Spatial Resolution.}
Figure~\ref{fig:q2-varydx} shows that obfuscating the spatial resolution of mobility events can also reduce the effectiveness of matching. More precisely, Figure~\ref{fig:q2-varydx} displays the probability of uniquely matching a trace, when fixing $\Delta_t$ to 5\,min, and varying the number of leaks together with $\Delta_{xy}$. Overall, we observe that higher spatial obfuscation decreases the probability of matching, as expected. Interestingly, we observe that setting $\Delta_{xy}=1$\,km does not provide sufficient protection for the traces, as the matching probability pattern is very similar to the case where $\Delta_{xy}= 0.2$\,km (see also Figure~\ref{fig:q1}). In fact, under this setting, 4 leaks are still sufficient to identify 92\% of the traces.
However, when $\Delta_{xy} = 5$\,km, we note that the lower spatial resolution reduces the performance of matching, since 8 leaks are required to match more than 90\% of the traces. Finally, with 10 leaks, 64\% of the traces are matched when $\Delta_{xy} = 25$\,km and 25\% when $\Delta_{xy}=125$\,km.

\begin{figure*}[t]
\center
    \begin{subfigure}[b]{0.22\textwidth}
		\includegraphics[width=1.0\textwidth]{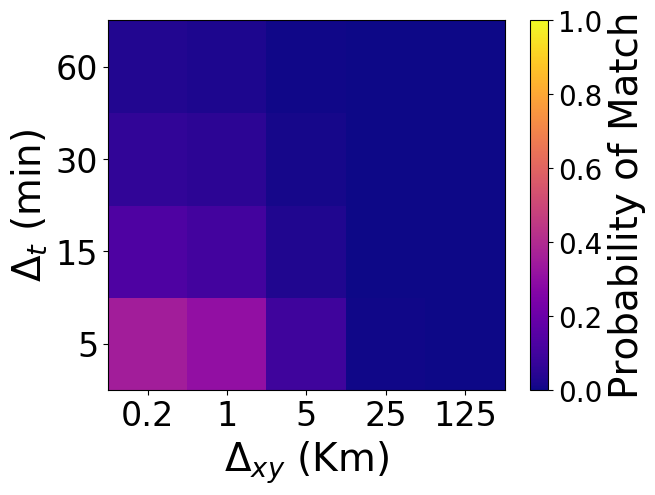}
		\caption{$k=2$}	
		\label{fig:q2-heat-k2}
	\end{subfigure}
	~
    \begin{subfigure}[b]{0.22\textwidth}
		\includegraphics[width=1.0\textwidth]{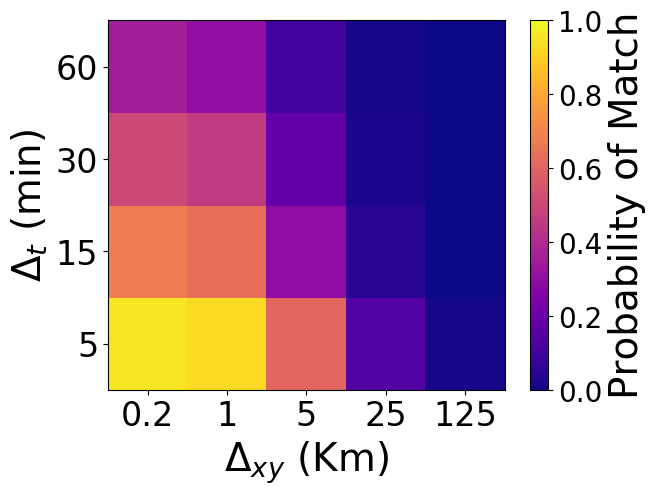}
        \caption{$k=4$}
        \label{fig:q2-heat-k4}
    \end{subfigure}
    ~
    \begin{subfigure}[b]{0.22\textwidth}
		\includegraphics[width=1.0\textwidth]{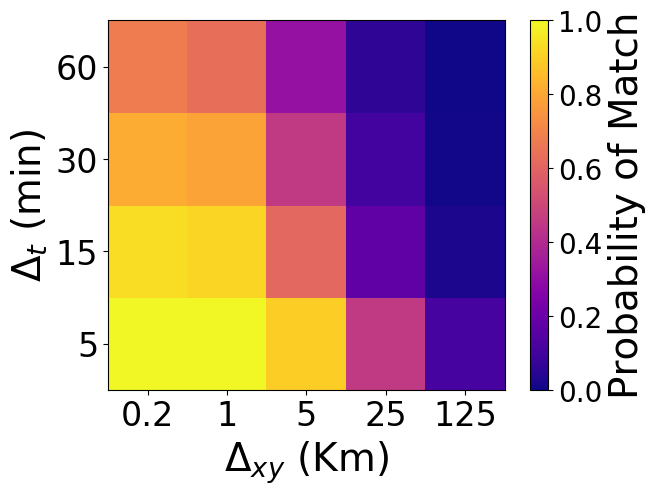}
        \caption{$k=7$}
        \label{fig:q2-heat-k7}
    \end{subfigure}
    ~
    \begin{subfigure}[b]{0.22\textwidth}
		\includegraphics[width=1.0\textwidth]{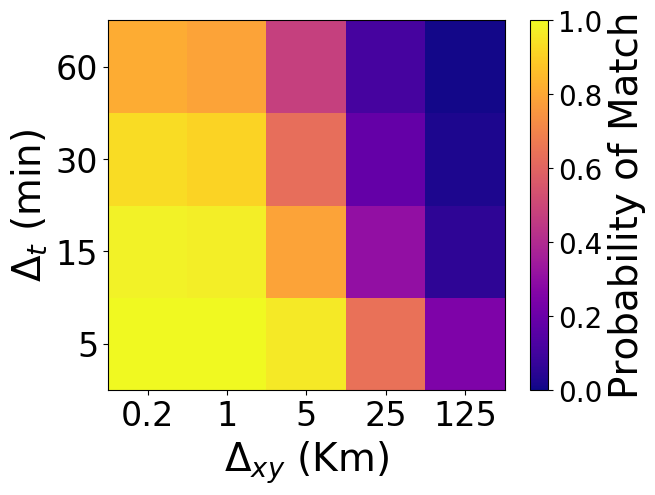}
        \caption{$k=10$}
        \label{fig:q2-heat-k10}
    \end{subfigure}
    \caption{Probability of uniquely matching a mobility trace vs. number of leaks ($k$) with variable $\Delta_t$ and $\Delta_{xy}$.}
    \label{fig:q2-heat}
    \vspace{-0.3cm}
\end{figure*}

\descr{Simultaneous Spatio-Temporal Obfuscation.}
Our previous set of experiments demonstrate how temporal or spatial obfuscation of the events contained in mobility traces can reduce the performance of matching.
We now attempt to understand the effect of the simultaneous application of both obfuscation approaches. Figure~\ref{fig:q2-heat} displays four heat maps depicting the probability of uniquely matching a trace for variable $\Delta_{t}$ and $\Delta_{xy}$ under different number of leaks $k \in \{2, 4, 7, 10\}$. Building on the previous results, and examining the corresponding axes of the heat maps, we observe that both types of obfuscation can reduce the probability of uniquely matching a trace, independently of $k$.

While comparing the colors of the heat maps at the top left (highest temporal obfuscation) and the bottom right (highest spatial obfuscation), we see that obfuscating the spatial dimension reduces the probability of finding a unique match more than just obfuscating the temporal one. This is not surprising, given that in our experimental setting the range of spatial granularity (from $0.2$ to $125$\,km) is much higher than the temporal one (from $5$ to $60$\,min). We also note that, as the number of leaks increases, the probability of matching also increases (the \textit{yellow} area increases from $k=2$ to $k=10$). For instance, when $k=4$, $\Delta_{xy}=1$\,km, and $\Delta_t=15$\,min, the probability of match is $0.63$, while when $k=10$ it goes up to $0.97$ for the same spatio-temporal granularity.

Looking at the top right of the heat maps, we see that performing simultaneous spatial and temporal obfuscation always brings better privacy protection than obfuscating just one of the dimensions (bottom right for spatial and top left for temporal). For instance, when 10 points are leaked from a trace, we observe that highly obfuscating the temporal dimension alone ($\Delta_t = 60$\,min) results to a matching probability of 0.8, while only obfuscating the spatial one ($\Delta_{xy}=125$\,km) reduces it to 0.25. Alternatively, when obfuscating both dimensions at the same time, the probability of finding a unique match goes down to 0.01. Nonetheless, we highlight the inherent trade-off between privacy and utility, as such levels of obfuscation potentially prohibit interesting applications on mobility data.

\descr{Takeaways:}\vspace{-0.1cm}
\begin{itemize}
    \item Lowering the temporal granularity from 5 to 60\,min (12$\times$ decrease) reduces the matching probability by 3$\times$ (considering 4 leaks and $\Delta_{xy}=0.2$\,km).
    \item Lowering the spatial granularity from 0.2 or 1\,km to 25\,km (25--125$\times$ decrease) reduces the matching probability by 4$\times$ (considering 4 leaks and $\Delta_{t}=5$\,min).
    \item 10 leaks with a temporal resolution of 60\,min are sufficient to match 81\% of users with $\Delta_{xy}=0.2$\,km.
    \item 10 leaks with a spatial resolution of 25\,km are sufficient to match 64\% of users with $\Delta_{t}=5$\,min.
    \item Performing simultaneous spatial and temporal obfuscation always brings better privacy protection than obfuscating just one of the dimensions.
\end{itemize}

\begin{figure}[t]
\centering
    \includegraphics[width=0.9\columnwidth]{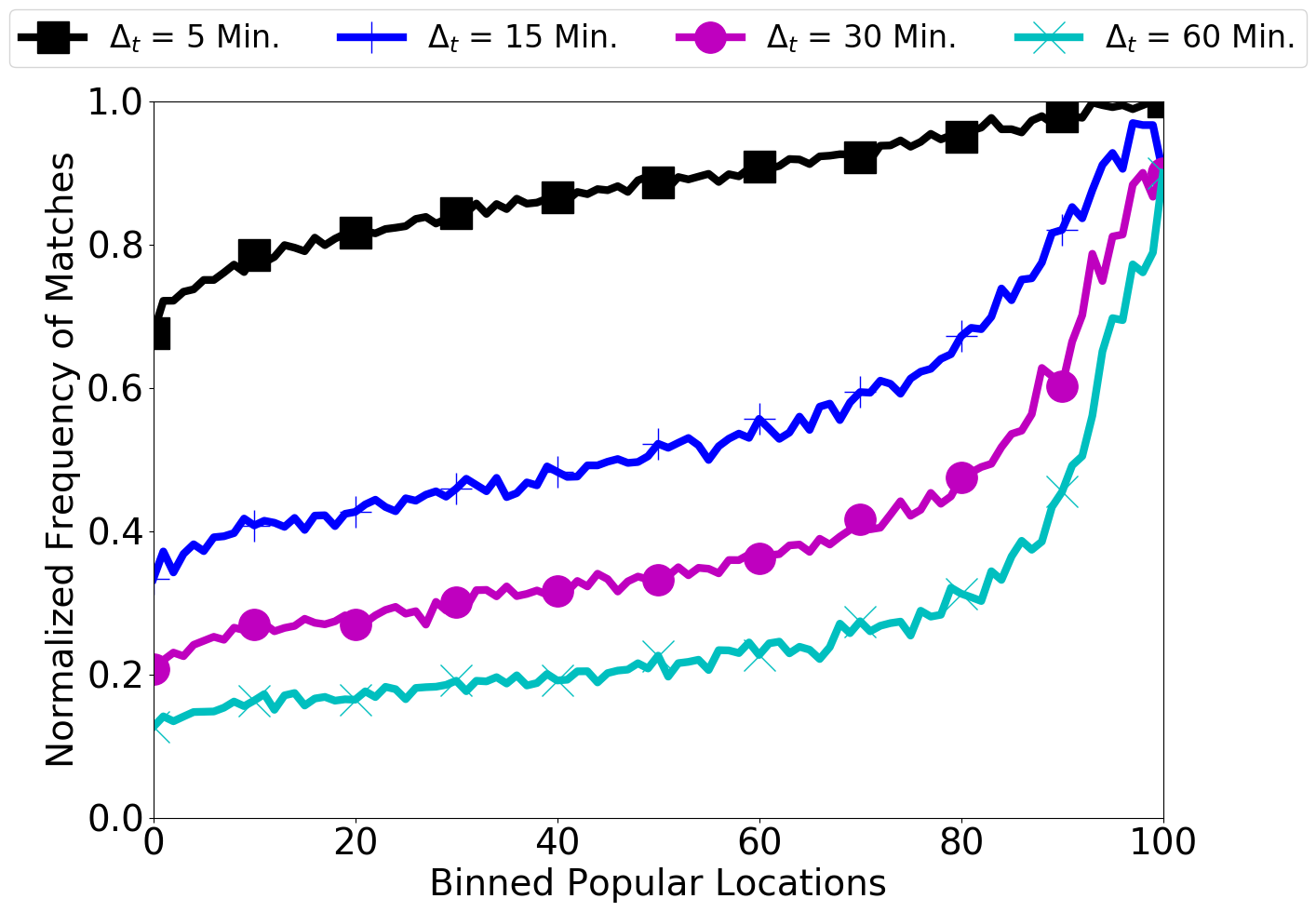}
    \caption{Normalized frequency of matches vs.~popular locations (in decreasing order), binned into 100 groups. Experiment conducted with $k=3$, $\Delta_{xy} = 0.2$\,km, and varying $\Delta_t$.}
    \label{fig:q4}
    \vspace{-0.1cm}
\end{figure}

\subsection{Does location popularity matter?}

We now examine which locations contribute to the unique matching of traces. To do so, we randomly sample $k=3$ events from 500\,k randomly selected traces and perform matching while we set $\Delta_{xy}=0.2$\,km and vary the temporal resolution. Next, we extract the locations of those spatio-temporal leaks that uniquely match a mobility trace and calculate their normalized frequency in the experiment, i.e.,~which percentage of times each location participated in a match.

Figure~\ref{fig:q4} displays the average normalized frequency of locations in the dataset, sorted by decreasing popularity (location's number of daily events) and grouped in 100 bins. We clearly observe that the frequency of matches increases as the leak corresponds to a less popular location. Therefore, this indicates that less visited locations are those that contribute more to the matching of a trace. This is expected as, intuitively, a trace of a mobile user moving in less popular (or less crowded) locations is harder to hide among the set of available traces.

In Figure~\ref{fig:q4}, we can also observe the effect of the temporal resolution on the matching. When $\Delta_t$ is fine-grained, e.g.,~5\,min, the most popular locations highly contribute to the identification of traces (0.7 avg.~normalized frequency), while when it is coarse-grained, e.g.,~60\,min, the normalized frequency decreases substantially (0.15). This shows that even crowded locations can help matching when there is sufficient detail on the temporal dimension. Alternatively, temporal generalization can mitigate this effect, as more users will appear in a location over the enforced time bins. Finally, we highlight that, irrespectively of the temporal resolution, the least popular locations (right part of the plot) allow for easy matching as their normalized frequency is very high.

\descr{Takeaways:}
\begin{itemize}
    \item Leaks from less popular locations are more likely matched to unique traces.
    \item Leaks from popular locations can still be matched to unique traces if the temporal resolution is fine-grained.
\end{itemize}

\begin{figure}[t]
\center
    \includegraphics[width=0.4\textwidth]{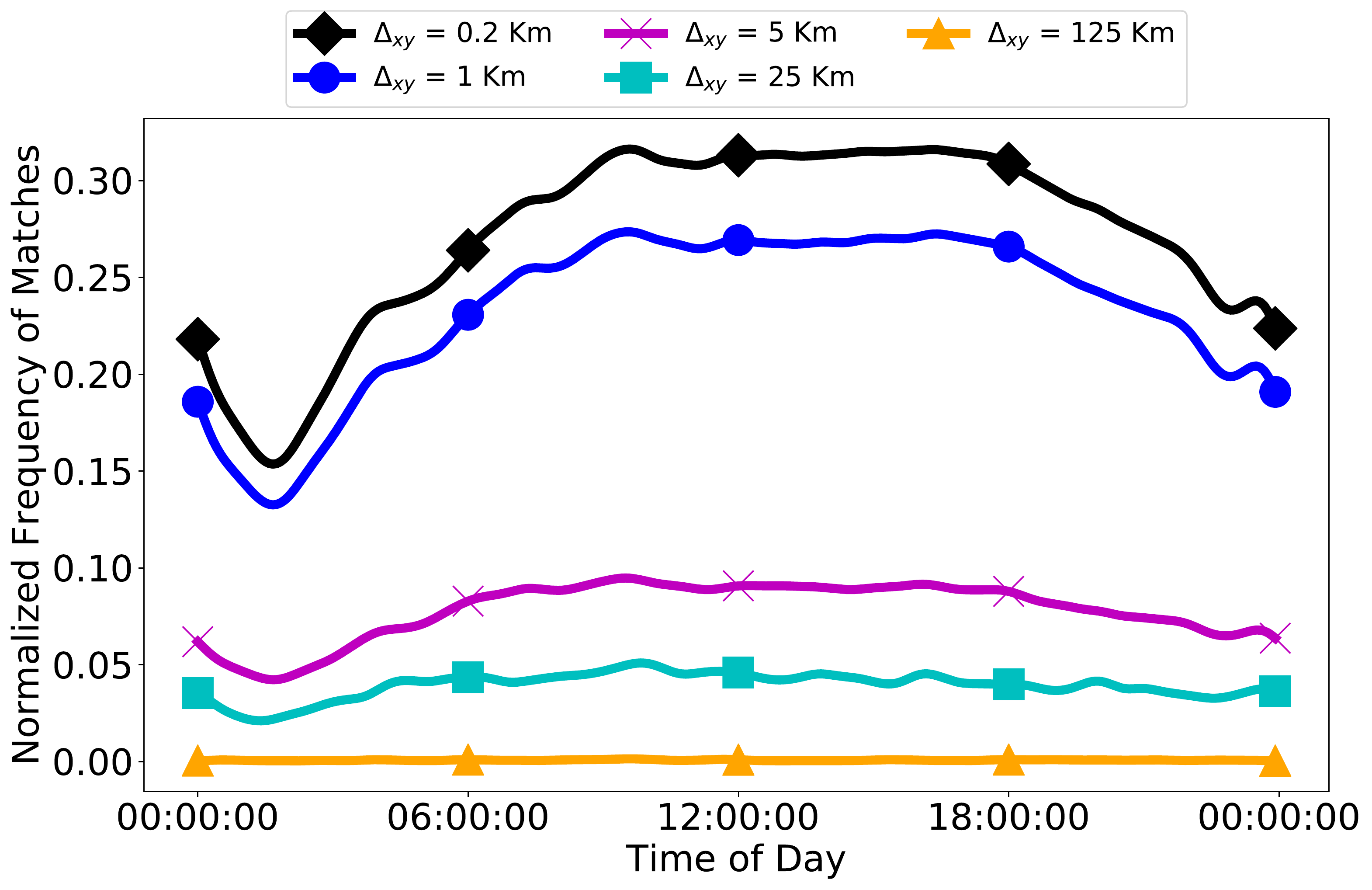}
    \vspace{-0.1cm}
    \caption{Normalized frequency of daily time slots. Experiment conducted with $k=3$, $\Delta_t = 5$\,min and $\Delta_{xy}=0.2$\,km.}
    \label{fig:q3}
    \vspace{-0.4cm}
\end{figure}

\subsection{Does time of day matter?}

Next, we investigate how the time of a spatio-temporal leak affects the performance of matching. We use the same sample used earlier, but fix $\Delta_t=5$\,min and vary $\Delta_{xy}$. Then, we extract the time slots of the spatio-temporal leaks that uniquely match a mobility trace and compute their normalized frequency. Figure~\ref{fig:q3} displays the corresponding results.

For fine-grained spatial resolution ($0.2$\,km), we note a clear diurnal behavior and a difference between day and night time. In particular, the plot shows that the contribution of time bins increases gradually in the early morning hours during commute time to work. During working hours (8:00--18:00) we observe the highest frequency, indicating that the high activity generated during this time frame contributes more to the identification of a trace. After being overall stable through the day, it decreases again in the evening hours. Moreover, it is clear that midnight hours (1:00--4:00) contribute less to the matching of traces, as people are stationary at their homes, or potentially turn off their phones during their sleep. Finally, we observe the effect of applying spatial generalization to the locations of leaks by looking at the different lines of the plot. While the frequency pattern is similar when $\Delta_{xy}=1$\,km, we note how further spatial obfuscation decreases the variability between time slots until, in the utmost case when $\Delta_{xy}=125$\,km, the different times of day do not affect the matching performance.

\descr{Takeaways:}
\begin{itemize}
    \item Location leaks generated while commuting or working contribute to de-anonymization more than leaks generated during night hours.
    \item Spatial obfuscation reduces the effect of time in the identification of a trace.
\end{itemize}

\begin{table}[t]
\center
    \resizebox{0.9\linewidth}{!}{
    \begin{tabular}{ | l || c | c | }
    \hline
    \textbf{Probability of match} & \textbf{1.0} & \textbf{ $<$ 0.01} \\ \hline \hline
    \textbf{Number of events} & 47.9 (30.1) & 29.9 (24.5) \\ \hline
    \textbf{Number of unique locations} & 16.7 (13.1) & 5.5 (5.6) \\ \hline
    \textbf{Area of bounding box (km$^2$)} & 550.5 (887.6) & 86.5 (360.8) \\ \hline
    \textbf{Distance traveled/time slot (km)} & 1.0 (1.7) & 0.2 (0.8) \\ \hline
    \textbf{Total distance traveled (km)} & 291.8 (490.1) & 62.5 (252.7) \\ \hline
    \textbf{Temporal entropy} & 2.6 (1.1) & 2.6 (1.1) \\ \hline
    \textbf{Spatial entropy} & 3.6 (0.7) & 3.1 (0.9) \\ \hline
    \end{tabular}
    }
    \caption{Mobility characteristics of traces vs.~probability of match. Experiment conducted with $k=3,$ $\Delta_t = 15$\,min, and $\Delta_{xy} = 1$\,km.
    Average (standard deviation) values are shown.}
    \label{table:q5}
    \vspace{-0.3cm}
\end{table}

\subsection{Do user mobility characteristics matter?}

In this experiment, we investigate the mobility characteristics of users whose traces can be uniquely matched using a handful of location leaks. To do so we use the previous sample, $k=3$, $\Delta_t=15$\,min, and $\Delta_{xy}=1$\,km. With that we examine those traces that are uniquely identified with probability 1. For comparison, we also show those traces that are matched with probability less than 0.01. From the traces of each case, we extract various statistics that capture general mobility characteristics:
(1) number of events $|M_u|$,
(2) number of unique locations in $M_u$,
(3) area of bounding box,
(4) distance traveled by the user per time slot,
(5) total distance traveled by the user, and
(6) temporal and spatial entropy.
The corresponding results are shown in Table~\ref{table:q5}.

Overall, the table shows that mobility traces with higher activity (e.g.,~because the user is moving a lot and/or using the mobile phone very often) are more likely to be matched using just very few spatio-temporal leaks. This is corroborated by observing the number of events in the traces, as traces that are uniquely matched contain on average 47.9 events compared to 29.9 for those which have a small chance of matching. Furthermore, we note that the uniquely matched traces move among a bigger set of locations (16.7 on average) in comparison to those matched with small probability (5.5). This indicates, once again, that high mobility eases user identification, as the large set of locations might correspond to users who are very active within cities (and where the antenna network is dense) or travel between cities using fast means of transport such as car and train. This last observation is further supported by examining the area of the traces' bounding box and the distances traveled.
Traces with high chance of matching move within larger geographic areas (550\,km$^2$), travel longer distances per time slot (1\,km on avgerage) and in total (291\,km). We can compare this with those with small chance of matching, who move within small areas (86\,km$^2$) and small travel distances (0.2\,km per time slot and 62\,km for the whole day). These results hint that the latter type of users mostly perform local movements within a city. Finally, while the temporal entropy does not show any significant correlation with the probability of matching a trace, the spatial one does. Traces with small chance of being matched have smaller entropy, indicating that users who move among smaller sets of locations probably match many other users and are difficult to uniquely identify.

\descr{Takeaway:}
\begin{itemize}
    \item Highly mobile users (either due to distance traveled, area covered, or both) are 100$\times$ more prone to identification.
\end{itemize}

\subsection{How unique is a user's mobility trace?}
\label{sec:results_uniqueness}

Thus far, our experiments demonstrated that a few spatio-temporal leaks are sufficient to identify a mobility trace. Now we investigate how our results relate to the uniqueness of full mobility traces. In other words, we look at how unique the daily mobility pattern of a user is. As discussed in our methodology (Section~\ref{sec:methodology}), we follow our matching approach to estimate both the upper and lower bounds on the probability of uniquely matching a trace, employing both \textit{strict} and more \textit{relaxed} matching criteria. For this last experiment, we use traces from a larger, randomly selected set of 1.2\,M users (again compared across all other users in our dataset) to compute the upper and lower bounds of uniqueness. We use the output of our matching algorithm to estimate the probability of obtaining a unique match in both cases. We repeat our experiments for variable $\Delta_t$ and $\Delta_{xy}$ to investigate the effect of the temporal and spatial obfuscation on the uniqueness of mobility traces.

\begin{figure}[t]
\center
    \includegraphics[width=0.9\columnwidth]{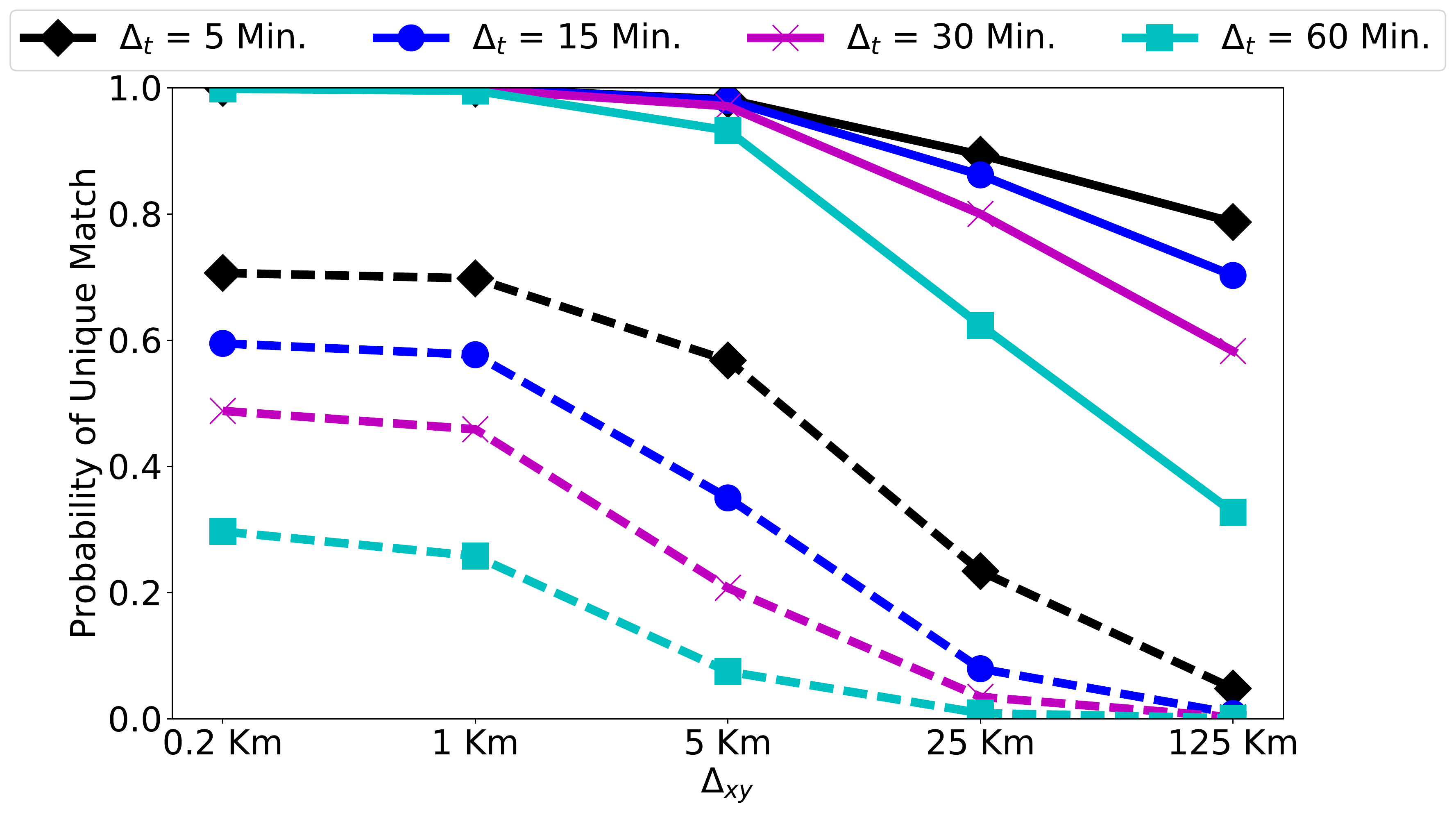}
    \vspace{-0.1cm}
    \caption{Probability of uniquely matching a mobility trace for variable $\Delta_t$ and $\Delta_{xy}$. Solid and dashed lines correspond to the estimated upper and lower bounds, respectively.}
    \label{fig:q6}
    \vspace{-0.5cm}
\end{figure}

Figure~\ref{fig:q6} displays the probability of uniquely matching a mobility trace with variable temporal and spatial resolution in the upper and lower bound cases. When we configure the criteria of our matching algorithm in a \textit{strict} manner, i.e.,~the upper bound case (solid lines), we note that the probability of matching a trace is very high. This is not surprising given our previous results, which showed that a handful of spatio-temporal leaks are sufficient to identify a trace. In particular, we highlight that for a spatial resolution of up to 5\,km, the probability of obtaining a unique match is very large ($\geq 0.9$), irrespectively of the temporal granularity. This indicates that large-scale mobility trajectories are extremely unique. When we further obfuscate trajectories in the space dimension, e.g.,~by setting $\Delta_{xy} = 25$\,km, we observe that traces become only slightly less distinct, as we increase the temporal granularity. For example, with $\Delta_t = 5$\,min, 90\% of traces are matched, compared to 60\% for $\Delta_t = 60$\,min. Nonetheless, we remark that mobility traces are still highly identifiable, as even in the case of extreme obfuscation in both space and time, i.e., with $\Delta_t = 60$\,min and $\Delta_{xy} = 125$\,km, the probability of obtaining a unique match reaches 0.32. This observation suggests that spatio-temporal obfuscation is not sufficient for the privacy protection of mobility traces.

Next, we \textit{relax} the criteria of our matching algorithm in an attempt to estimate a lower bound on the probability of uniquely matching a trace. The dashed lines of Figure~\ref{fig:q6} display the corresponding results for variable temporal and spatial resolutions. While we now observe that, overall, the probability of matching is reduced (compared to the upper bound), it is surprising that it remains high for fine-grained temporal and spatial resolutions. For instance, when $\Delta_t=5$\,min and $\Delta_{xy}=0.2$\,km, 70\% of the traces remain unique despite our loose matching criteria. Moreover, we remark how spatio-temporal obfuscation reduces faster the effectiveness of matching when compared to the upper bound case. With $\Delta_t=15$\,min and $\Delta_{xy}=0.2$\,km, there is a 60\% chance of uniquely matching a trajectory. Furthermore, the probability goes down to 7\% and to 1\% when $\Delta_{xy}=25$\,km and $125$\,km, respectively. Note how the same setting in the upper bound case achieves smaller reduction (ranging from a probability of $1$ down to $0.75$).

Generalizing trajectories in time also makes them less identifiable, as we can realize by comparing the different color lines of the lower part of the figure. As an example, when $\Delta_{xy} = 1$\,km, the probability of matching a trace is $0.7$ with $\Delta_t=5$\,min, while it goes down to $0.3$ when $\Delta_t=60$\,min. Finally, we remark how simultaneous temporal and spatial obfuscation reduces significantly the probability of matching, e.g., in the extreme case where $\Delta_t = 60$\,min and $\Delta_{xy} = 125$\,km, it almost reaches zero. This hints that when spatio-temporal obfuscation is combined with other techniques like suppression (hiding events that uniquely identify a user) and noise injection (introduce \textit{fake} events to favor matches) is more effective for the protection of mobility traces.

\descr{Takeaways:}
\begin{itemize}
    \item For spatial resolution of up to 5\,km, the matching probability is very large ($\geq 0.9$), irrespectively of the temporal granularity.
    \item Even with extreme obfuscation in both space and time ($\Delta_t = 60$\,min, $\Delta_{xy} = 125$\,km), the matching probability remains considerably high ($0.32$).
\end{itemize}
\section{Discussion}
\label{sec:discussion}

To the best of our knowledge, this is the first study on privacy and mobility that uses a really large (tens of millions of users) and, at the same time, fine-grained dataset (two orders of magnitude more location events than CDRs).

\subsection{Main findings} 

Our analysis shows that, even among tens of millions of users, a handful of anonymous location leaks are sufficient to identify a user, based on her stored mobility trace. In fact, only one leak is enough to exclude 99.999\% of the candidate traces, and 3 (resp.~4) anonymous location leaks are sufficient to uniquely match 80\% (resp.~95\%) of the traces.

In terms of why some users are more identifiable than others, we observe that leaks from less popular locations and leaks during commuting and working hours greatly harm users' privacy. Unsurprisingly, highly mobile users are 100$\times$ more prone to identification, especially users covering large geographic areas, traveling long distances, and visiting a diverse set of locations during their movements.

We also show that, in general, lowering the temporal and spatial granularity of leaks significantly impairs identification. For example, obfuscating the temporal (resp.~spatial) dimension by 12$\times$ (resp.~25--125$\times$), can reduce the identifiability by 3$\times$ (resp.~4$\times$). Performing simultaneous obfuscation in both dimensions brings better privacy protection than applying one of the two alone: reducing the two dimensions to the lowest considered granularity ($\Delta_t = 60$\,min and $\Delta_{xy} = 125$\,km) reduces the probability of matching by 100$\times$. Finally, while spatio-temporal obfuscation reduces the probability of matching whole mobility traces between themselves, it is not alone sufficient to mitigate the privacy threat represented by the uniqueness of large-scale mobility traces.

\subsection{Limitations} 

Our measurement study only uses one day worth of data. Extending the tracking period for weeks or even months can only make users even more identifiable. Furthermore, to protect the privacy of our users, we did not attempt to de-anonymize them based on data originating from different platforms. Instead, we only sample events from our own logs, and create artificial location leaks. As the main focus of our study is to test the uniqueness of mobility patterns and the feasibility of user re-identification, we did not address the issue of having leaks that do not exist in the fine-grained logs. Finally, our results are tailored to the mobility behavior of the population of the considered country over a working day. Replicating our results over weekends or other seasons, or with a focus on countries with different mobility behaviors is part of our future work.

\subsection{Implications}

While some organizations that manage data for operational purposes (ISPs, governments, insurance companies) are heavily regulated and are likely to respect users' privacy, many other companies are either unregulated or may choose to ignore privacy laws imposed on a certain region (for instance the EU~\cite{link2} and specifically regarding the use of personal data and the GDPR~\cite{gdpr}).
Such companies are typically Internet-based, and depend on the availability and exploitation of data in order to generate revenues. For example, they can be involved in the aggregation of online users' personal data, including web preferences and locations, in order to offer better, ``free'' services~\cite{www18adcost}. However, in exchange, and not explicitly communicated to the user, they sell collected user data to interested third-parties. This is typically the case within the real-time bidding ad-ecosystem, where data such as location and user behavioral traits are sold for better ad-targeting~\cite{imcRTB}.

Similarly, mobile app developers frequently collect and share location information with third-party companies, in order to perform better ad-targeting. Additionally, many online publishers host on their websites third-party trackers that collect and cross-reference location data (based on user IPs), for targeting or personalization purposes. In all these scenarios, typical mechanisms employed such as cookie synchronization have already been shown to leak information like browsing visits and location to tens of colluding parties, even if the user employs virtual private networks and SSL sessions to hide her location~\cite{eurosec18csyncssl}.

Our results show that it is possible for large or small online entities to easily identify a user with a few location leaks. But identification is only one part of the story, as identities may be further linked to a myriad of other types of information, and obviously the rest of the places the user has been become exposed, with all the implications that this may have. In summary, we show that it is possible for various companies to identify a user or her mobility trace across different sets of data they have either bought from some data management platform, or have been shared with them for better targeting, or have been accidentally leaked by the user and collected by some mobile app or website.

\subsection{What can be done?} 

Our work demonstrates the need for location obfuscation techniques by the entities that maintain large-scale mobility logs. Our results should be taken into account by service providers that desire to bring privacy protection to the mobility traces they store or publish. Using similar settings as those in our experiments, a service provider can estimate the amount of temporal and spatial obfuscation required to achieve a sufficient protection as measured by probability of uniquely matching a trace, given a certain number of leaks. Such obfuscation can then be applied on the provider's trace logs to guarantee users' privacy and compliance with privacy laws.

Other mitigation measures include the addition of noise in the storage of trace logs and the obfuscation of the exact location of users while still maintaining the utility of the data~\cite{quercia11spotme}. Moreover, location events could be suppressed from a trace to make it less identifiable. In fact, the service provider could perform storage and retrieval of user location data in an adaptive manner (i.e.,~by tuning the rate of event storage or retrieval) to provide guarantees of user anonymity.

Finally, online platform providers such as Facebook, Twitter, Google, or Foursquare, assuming they are serious about supporting user privacy, they could proactively warn users when their next post(s) or search(es) would lead to possible re-identification by third-party trackers, based on the users' historical mobility trace. As an active obfuscation measure from these platforms, noise could be added by delaying the post to appear on the online page or by providing a less accurate geo-location.
\section{Conclusion}
\label{sec:conclusion}

In this work, we leverage an unprecedented scale, country-wide dataset of network events from an European mobile operator to study the extent to which anonymous location leaks can be uniquely mapped to a mobility trace and its corresponding identity. While our results demonstrate that limited location information within the time frame of a day is sufficient to uniquely identify a mobility trace, we also expose the main contributing factors to de-anonymization. Furthermore, we show that while spatio-temporal obfuscation helps at reducing the probability of de-anonymization, alone, it is not sufficient to mitigate the privacy threat represented by the uniqueness of large-scale mobility traces. Nonetheless, the insights gained from our measurement study can be used as guidelines for the design of novel privacy protection mechanisms or policies regarding the storage and publication of mobility data.

\section*{Acknowledgment}
This work was partially supported by the I-BiDaaS project, funded by European Commission under Grant Agreement No. 780787.
This publication reflects the views only of the authors, and the Commission cannot be held responsible for any use which may be made of the information contained therein.

\bibliographystyle{abbrv}
\bibliography{main}

\end{document}